%Paper: hep-th/9409044
%From: gregory moore <moore@castalia.physics.yale.edu>
%Date: Thu, 8 Sep 94 23:26:37 -0400

\input harvmac.tex

\def\IL{\relax{\rm I\kern-.18em L}}
\def\IQ{\relax\hbox{$\inbar\kern-.3em{\rm Q}$}}
\def\IH{\relax{\rm I\kern-.18em H}}
\def\IR{\relax{\rm I\kern-.18em R}}
\def\IC{\relax\hbox{$\inbar\kern-.3em{\rm C}$}}
\def\IZ{\relax\ifmmode\mathchoice
{\hbox{\cmss Z\kern-.4em Z}}{\hbox{\cmss Z\kern-.4em Z}}
{\lower.9pt\hbox{\cmsss Z\kern-.4em Z}}
{\lower1.2pt\hbox{\cmsss Z\kern-.4em Z}}\else{\cmss Z\kern-.4em Z}\fi}
\def\CM {{\cal M}}
\def\ST{{\Sigma_T}}
\def\Sw{{\Sigma_w}}

\def\CG {{\cal G}}
\def\CQ {{\cal Q}}

\def\CF {{\cal F}}

\def\CL {{\cal L}}
\def\CV {{\cal V}}
\def\CO {{\cal O}}
\def\CZ {{\cal Z}}
\def\CE {{\cal E}}
\def\CH {{\cal H}}
\def\CC {{\cal C}}

\def\CA{{\cal A}}
\def\CY{{\cal Y}}

\def\O{\Omega}
\def\ha{{1\over 2}}

\def\lieg{g}
\def\liebg{Lie(\CG)}
\def\ymt{${YM_2}$}
\def\Tr{\rm Tr}

\def\dim{\rm dim}
\def\cok{\rm cok}
\def\exp{\rm exp}
\def\vol{\rm vol}
\def\lieg{{\underline{\bf g}}}
\def\inbar{\,\vrule height1.5ex width.4pt depth0pt}
\def\IF{\relax{\rm I\kern-.18em F}}
\def\IO{\relax\hbox{$\inbar\kern-.3em{\rm O}$}}
\font\cmss=cmss10 \font\cmsss=cmss10 at 7pt
\def\IZ{\relax\ifmmode\mathchoice
{\hbox{\cmss Z\kern-.4em Z}}{\hbox{\cmss Z\kern-.4em Z}}
{\lower.9pt\hbox{\cmsss Z\kern-.4em Z}}
{\lower1.2pt\hbox{\cmsss Z\kern-.4em Z}}\else{\cmss Z\kern-.4em
Z}\fi}
\def\SDiff{{\rm SDiff}}
\def\Diff{{\rm Diff}}
\def\Met{{\rm Met}}
\def\Weyl{{\rm Weyl}}
\def\Hol{{\rm Hol}}
\def\Map{{\rm Map}}
\let\lref=\nref
\lref\AsMo{P.~ Aspinwall and D.~ Morrison, ``Topological Field Theory
and Rational Curves'', Commun. Math. Phys. {\bf 151} (1993) 245,
hep-th/9110048.}

\lref\AtBoym{M. F. Atiyah and R. Bott, ``The Yang-Mills Equations Over Riemann
Surfaces,'' Phil. Trans. Roy. Soc. London {\bf A308} (1982) 523.}
\lref\AtJe{M.F.Atiyah and L.Jeffrey, ``Topological Lagrangians and
Cohomology,'' Jour. Geom.  Phys. {\bf 7} (1990) 119.}
\lref\atiyahsinger{M.F. Atiyah and I.M. Singer,
``Dirac operators coupled to vector potentials,''
Proc. Natl. Acad. Sci. {\bf 81}(1984)2597}
\lref\audin{M. Audin and J. Lafontaine, {\it Holomorphic
curves in symplectic geometry} Birkh\"auser, 1994}
\lref\BaSi{L.Baulieu and I.Singer, ``The Topological
sigma model,'' Commun. Math. Phys. 125, 227-237 (1989).  }
\lref\BaSiii{L.Baulieu and I.Singer, ``Topological
Yang-Mills Symmetry,'' Nucl. Phys. B. Proc. Suppl.
5B (1988)12}
\lref\bbrt{Birmingham, Blau, Rakowski, and Thompson,
``Topological Field Theories,'' Phys. Reports {\bf 209}(1991)129. }
\lref\Bl{M. Blau, ``The Mathai-Quillen Formalism and Topological
Field Theory,'' Notes of Lectures given at the Karpacz Winter School on
`Infinite Dimensional Geometry in Physics', hep-th/9203026 }
\lref\CMROLD{S. Cordes, G. Moore, and S. Ramgoolam,
``Large N 2D Yang-Mills Theory and Topological String Theory,''
hep-th/9402107}
\lref\cmrrev{S. Cordes, G. Moore,
and  S. Ramgoolam, ``Lectures on 2D Yang-Mills theory,
equivariant cohomology and topological
field theory, ''  Yale preprint,  YCTP-P11-94. }
\lref\DiVeVe{R. Dijkgraaf, E. Verlinde and H. Verlinde Nucl. Phys. {\bf B348}
(1991) 435; {\bf B352} (1991) 59; in String Theory and quantum
Gravity, Proc. Trieste Spring School, April 1990 (World Scientific,
Singapore, 1991).}
\lref\Dosc{M. R. Douglas, ``Some Comments on QCD String,''
To appear in Proceedings of the Strings '93 Berkeley conference.}
\lref\Docft{M.R. Douglas, ``Conformal Field Theory Techniques
in Large $N$ Yang-Mills Theory,''
hep-th/9311130,to be published in the proceedings of the May 1993 Carg\`ese
workshop on Strings,
Conformal Models and Topological Field Theories. }
\lref\DoKa{M.~ Douglas and Kazakov,
``Large $N$ Phase Transition in Continuum QCD$_2$''
hep-th/9305047 }
\lref\FMS{D. Friedan, E. Martinec, and S. Shenker, Nucl. Phys. B271 (1986) 93.}
\lref\Fulton{W. Fulton, Hurwitz schemes and
irreducibility
of moduli of algebraic curves, Annals of Math. 90, 542 (1969). }\lref\GrTa{
D. Gross, W. Taylor, `Two-dimensional QCD is a String Theory,'
hep-th/9301068;
D. Gross, W. Taylor, `Twists and Loops in the String
Theory of Two Dimensional QCD,'
hep-th/9303046.}
\lref\GrTatalk{D.J. Gross and W. Taylor, ``Two-Dimensional QCD
and Strings,'' Talk presented at Strings '93, Berkeley, 1993;
hep-th/9311072}.
\lref\Grtalk{D. Gross, ``Some new/old approaches to QCD,''
hep-th/9212148; ``Two Dimensional
QCD as a String Theory,''  hep-th/9212149,
Nucl. Phys. B400 (1993) 161-180}
\lref\grssmatyt{D. Gross and A. Matytsin, ``Instanton Induced Large $N$ Phase
Transitions in Two and Four Dimensional QCD,''
hep-th/9404004}
\lref\Hora{P. Horava, ``Topological Strings and QCD in Two Dimensions,''
EFI-93-66, hep-th/9311156. To appear in Proc. of The Cargese Workshop,
1993. }
\lref\jeffkir{L. Jeffrey and F. Kirwan, ``Localization for Nonabelian
Group Actions,''  alg-geom/9307001.}
\lref\kcpt{V.G. Kac and D.H. Peterson,
``Infinite dimensional
Lie algebras, theta-functions, and modular forms,''
Adv. in Math. {\bf 53} (1984)125}
\lref\Kal{J.~ Kalkman, ``BRST Model for Equivariant Cohomology
and Representatives for the Equivariant Thom Class",Commun. Math.
Phys. {\bf 153} (1993) 447;``BRST model applied to symplectic
geometry,'' hep-th/9308132}
\lref\Ki{F. Kirwan, {\it Cohomology of Quotients In Symplectic
And Algebraic Geometry,} Princeton University Press.}
\lref\LaPeWi{ J.M.F Labastida, M. Pernici,and E. Witten ``Topological
Gravity in Two dimensions,'' Nucl. Phys. B310,611}
\lref\MaQu{V.~ Mathai and D.~ Quillen,
``Superconnections, Thom Classes, and Equivariant Differential Forms",
Topology {\bf 25} (1986) 85.}
\lref\Mig{A. Migdal, Zh. Eksp. Teor. Fiz. {\bf 69} (1975) 810
(Sov. Phys. Jetp. {\bf 42} 413).}
\lref\MitH{A. A. Migdal,
 `` Loop equations and 1/N expansion'',
Phys. Reports {\bf 102},
(1983)199-290}
\lref\MiPoetd{J.~ Minahan and A.~ Polychronakos, "Equivalence of Two
Dimensional QCD and the $c=1$ Matrix Model'',  hep-th/9303153.}
\lref\mrsb{G. Moore and N. Seiberg,
 ``Polynomial Equations for Rational Conformal Field
Theories,''
Phys. Lett.  {\bf 212B}(1988)451;
``Classical and Quantum Conformal Field Theory,''
Commun. Math. Phys. {\bf 123}(1989)177;
``Lectures on Rational Conformal Field Theory,''
 in {\it Strings 90}, the proceedings
of the 1990 Trieste Spring School on Superstrings.}
\lref\stora{S. Ouvry, R. Stora, and P. Van Baal,
``On the Algebraic Characterization of Witten's
Topological Yang-Mills Theory,''
Phys. Lett. {\bf 220B}(1989)159}
\lref\zuckerman{The following construction
goes back to
D. Quillen, ``Rational homotopy theory,'' Ann. of
Math. {\bf 90}(1969)205. The application to the
present case was developed with G. Zuckerman.}
\lref\tH{G.~ `t Hooft,
``A Planar Diagram Theory for Strong
Interactions,
Nucl. Phys. {\bf B72} (1974) 461.}
\lref\verlinde{E. Verlinde,
``Fusion Rules and Modular Transformations
in 2d Conformal Field Theory,''
 Nucl. Phys. {\bf B300}(1988)360}
\lref\VeVe{E. Verlinde and H. Verlinde , Nucl. Phys.
{\bf B348} (1991) 457.}
\lref\Wi{K.G. Wilson, Phys. Rev. {\bf D10},2445 (1974).}
\lref\Witdgt{ E. Witten, ``On Quantum gauge theories in two dimensions,''
Commun. Math. Phys. 141, 153 (1991).}
\lref\Witdgtr{E.~ Witten, ``Two Dimensional Gauge Theories
Revisited", hep-th/9204083, J. Phys. {\bf G9} (1992) 303}
\lref\cohoft{E. Witten, Introduction to Cohomological
Field Theory,  in Trieste Quantum Field Theory 1990:15-32 (QC174.45:C63:1990)}
\lref\Winm{E.~ Witten, ``The N-Matrix Model
and gauged WZW models", Nucl. Phys. {\bf B371} (1992)
191.}
\lref\wttnmirror{E. Witten, ``Mirror Manifolds
And Topological Field Theory,''
hep-th/9112056, in {\it Essays on Mirror Manifolds} International
Press 1992}
\lref\Witp{E. Witten,
``On the structure of the topological phase
of Two dimensional Gravity,''
Nucl. Phys. B340 (1990) 281}
\lref\donaldson{E. Witten, Topological Quantum Field
Theory, Commun. Math. Phys. {\bf 117}(1988)353}
\lref\Witsm{E. Witten, ``Topological sigma models '', Commun.
Math. Phys. 118, 411-419 (1988). }

\def\bO{\IO}
% Macros for boxes

\def\boxit#1{\vbox{\hrule\hbox{\vrule\kern8pt
\vbox{\hbox{\kern8pt}\hbox{\vbox{#1}}\hbox{\kern8pt}}
\kern8pt\vrule}\hrule}}
\def\mathboxit#1{\vbox{\hrule\hbox{\vrule\kern8pt\vbox{\kern8pt
\hbox{$\displaystyle #1$}\kern8pt}\kern8pt\vrule}\hrule}}

\font\manual=manfnt \def\dbend{\lower3.5pt\hbox{\manual\char127}}

\def\IZ{\relax\ifmmode\mathchoice
{\hbox{\cmss Z\kern-.4em Z}}{\hbox{\cmss Z\kern-.4em Z}}
{\lower.9pt\hbox{\cmsss Z\kern-.4em Z}}
{\lower1.2pt\hbox{\cmsss Z\kern-.4em Z}}\else{\cmss Z\kern-.4em Z}\fi}
\def\half {{1\over 2}}
\def\sdtimes{\mathbin{\hbox{\hskip2pt\vrule height 4.1pt depth -.3pt width
.25pt
\hskip-2pt$\times$}}}
\def\p{\partial}

\Title{ \vbox{\baselineskip12pt\hbox{hep-th/9409044}
\hbox{YCTP-P10-94}}}
{\vbox{
\centerline{2D Yang-Mills Theory }
\centerline{and}
\centerline{Topological Field Theory}}}
\bigskip
\centerline{Gregory Moore}
\bigskip
\centerline{moore@castalia.physics.yale.edu}
\smallskip\centerline{Dept. of Physics, Yale University,New Haven, CT}
\bigskip

Contribution to the Proceedings of the International
Congress of Mathematicians 1994.
We review recent developments in the physics
and mathematics of Yang-Mills
theory in two dimensional spacetimes.

\Date{August 2,1994; Revised, Sep. 8, 1994 }
%\draft

\newsec{Introduction}

Two-dimensional Yang-Mills theory (\ymt) is
often dismissed as a trivial system. In fact it is
very rich mathematically and might be the source of
some important lessons physically.

Mathematically
\ymt\ has served as a
tool for the study of the topology of the moduli
spaces of flat connections on surfaces
\refs{\AtBoym,\Ki,\Witdgt,\Witdgtr}. Moreover,
recent work has shown that it contains much information
about the topology of Hurwitz spaces - moduli spaces
of coverings of surfaces by surfaces.

Physically,
\ymt\ is important because it is the first example
of a nonabelian gauge theory which can be reformulated
as a string theory. Such a reformulation
 offers one of the few ways in
which analytic results could be obtained for
strongly coupled gauge theories.
Motivations for a string reformulation
include experimental ``approximate duality''
of strong interaction amplitudes,
weak coupling expansions \tH,
strong coupling expansions \Wi\
and loop equations \MitH.
The evidence is suggestive but far from conclusive.
In \Grtalk\  D. Gross proposed the search for a string
formulation
of Yang-Mills theory using the exact results of
\ymt. This program has enjoyed some
success. A successful outcome for $YM_4$ would
have profound consequences, both mathematical and
physical.

In order to describe
the string interpretation of \ymt\
properly we will be led
to a subject of broader significance:
the construction
of cohomological field theory (CohFT).
This is reviewed in section 6.

\newsec{ Exact Solution of \ymt }

Let $\ST$ be a closed 2-surface equipped with
Euclidean metric.  Let
$G$ be a compact
Lie group with Lie algebra $\lieg$,
$P\to \ST$ a principal $G$-bundle,
$\CG(P)=Aut(P)$,
$\CA(P)=$  the space of connections on $P$.
The action for \ymt\  is the $\CG(P)$-invariant
function on $\CA(P)$ defined by:
$I_{\rm YM} = {1\over 4 e^2}\int_{\ST}
\Tr (F\wedge * F)$; $F=dA+A^2$,
$*=$ Hodge dual,  $e^2=$ gauge coupling.
$I_{\rm YM}$ is equivalent to a theory with action:
$I(\phi,A)= -\half \int_\ST  i \Tr (\phi F)  + \half e^2 \mu  \Tr \phi^2 $;
$\phi\in \O^0(M;\lieg)$, $\mu=*1$, and $\Tr$ is normalized
as in \Witdgt: ${1\over 8 \pi^2} \Tr F^2$ represents
the fundamental class of $H^4(B\tilde G;\IZ)$, where
$\tilde G$ is the universal cover of $G$.
Various definitions of the quantum theory
will differ by a renormalization ambiguity
$\Delta I = \alpha_1 \int {R\over 4 \pi}  + \alpha_2 e^2 \int \mu$.
Equivalence to the theory $I(\phi,A)$ shows that
\ymt\  is $\SDiff(\ST)$ invariant (no gluons!)
and that amplitudes are functions only of
the topology of $\ST$ and  $e^2 a$,  where $a=\int \mu$.

The Hilbert space $\CH_G$ is the space of class functions
$L^2(G)^{Ad(G)}$ and has a natural basis given by
unitary irreps:
$\CH_{G} = \oplus_R \IC\cdot \mid R\rangle$.
The Hamiltonian is essentially the
quadratic Casimir: $C_2 + \alpha_2$.
The amplitudes are nicely summarized using standard
ideas from topological field theory. Let
 ${\bf \underline{S}}$ be the tensor category of
oriented surfaces with area: $Obj({\bf \underline{S}})$=
disjoint oriented circles, $Mor({\bf \underline{S}})$=
oriented cobordisms, then:

\noindent
{\bf Theorem 2.1}: \ymt\ amplitudes provide a representation
of the geometric category ${\bf \underline{S}}$.
The state associated to the cap of area $a$ is:
$$e^{\alpha_1}
\sum_R \dim R e^{-e^2 a (C_2(R)+\alpha_2) } \mid R\rangle \qquad. $$
The morphism associated to the tube is
$$  \sum_R e^{-e^2 a  (C_2(R)+\alpha_2)}
 \mid R\rangle \langle R\mid \quad ,
$$
and the trinion with
two ingoing and one outgoing circle is:
$$ e^{-\alpha_1} \sum_R (\dim R)^{-1} e^{- e^2a( C_2(R)+\alpha_2)}
 \mid R\rangle \langle R\mid\otimes \langle R\mid
\qquad .
$$

\noindent
{\it Proof}: The heat kernel defines
a renormalization-group invariant plaquette
action $\spadesuit$

\noindent
{\bf Corollary:} On a closed oriented
surface $\ST$ of area $a$ and
genus $p$ the partition function is
\eqn\prfnct{
Z(e^2a, p,G ) = e^{\alpha_1 (2-2p)}
\sum_R (\dim R)^{2-2p} e^{- e^2 a (C_2(R)+ \alpha_2) }
}
These considerations
go back to \Mig.   A  clear exposition
is in \Witdgt.

\newsec{\ymt\ and the moduli space of flat bundles }

At $e^2 a=0$ the action $I(\phi,A)$ defines a
topological field theory ``of Schwarz type''
\bbrt.  In \Witdgt\Witdgtr\
Witten applied \ymt\ to the study of the topology of the
space of flat $G$-connections on $\ST$:
$\CM \equiv \CM(F=0; \ST, P) =
\{ A\in \CA(P): F(A) =0\}/\CG(P)$.
\foot{We take a
topologically trivial $P$ for simplicity.
$\CM$ then has singularities, but the results extend
to the case of twisted $P$, where  $\CM$
can be smooth \AtBoym.}

Witten's first result is that, for appropriate choice of
$\alpha_1$,  $Z$ computes the symplectic
volume of $\CM$ \Witdgt:
\eqn\symvol{
Z(0,p,G) = {1\over \# Z(G) } \int_\CM \exp \omega
={1\over \# Z(G) } \vol(\CM)}
 where $Z(G)$ is the center, and
$\omega$  is the symplectic form on $\CM$ inherited from
the 2-form on $\CA$:
$\omega(\delta A_1 , \delta A_2) = {1\over 4 \pi^2}
\int_\Sigma \Tr ( \delta A_1\wedge \delta A_2)$.
The argument uses a careful application of
Faddeev-Popov  gauge fixing and the triviality
of analytic torsion on oriented two -surfaces.
The result extends to the unorientable case, and
the constant $\alpha_1$ can be evaluated by
a direct computation of the Reidemeister torsion.

According to \Witdgt,  \symvol\  is  the
large $k$ limit of  the Verlinde  formula \verlinde.
Let $S_{RR'}(k)$ be the modular transformation matrix
for  the characters
of integrable highest weight modules $R\in P_+^k$
of the affine Lie algebra $\lieg_k^{(1)}$
under $\tau\to -1/\tau$ \kcpt.
At $e^2 a=0$ we have:
 $Z=\lim_{k\to \infty} e^{\alpha_1 \chi(\ST)} \sum_{R\in P_+^k}
({S_{00}(k)\over S_{0R}(k)})^{2p-2}$ where
$0$ denotes the basic representation. On the other hand,
 we may choose a complex structure $J$ on $\ST$
inducing a holomorphic line bundle
$\CL\to \CM$  with $c_1(\CL)=\omega$, and apply the
Verlinde formula to get:  $\lim_{k\to\infty} k^{-n}
\sum_{P_+^k}  ({1\over S_{0R}})^{2p-2} =
\lim_{k\to \infty} k^{-n} \dim  H^0(\ST;\CL^{\otimes k})
= \lim_{k\to \infty} k^{-n} \langle e^{k c_1(\CL)} Td\CM, \CM\rangle
= \vol \CM $, where $n=\half \dim \CM$.
Using \kcpt\  one recovers \symvol\ with
$$e^{\alpha_1} = (2 \pi)^{\dim G} /(\sqrt{\mid P/L\mid} \vol G)
=(\prod_{\alpha>0} 2 \pi (\alpha,\rho))/ \sqrt{\mid P/L\mid} \quad ,
$$
 $P$ is the weight lattice, $L$ the long root lattice,
and $\rho$ the Weyl vector.
The fact that the trinion is
diagonal in the sum over representations is the
 large $k$ limit of
Verlinde's diagonalization of fusion rules.
\foot{first proved, using conformal field
theoretic techniques, in \mrsb.}

Witten's second result \Witdgtr\  gives the
asymptotics of \prfnct\ for  $e^2a\to 0$ (set $a=1$):
\eqn\smllcpl{
Z(e^2, p,G) ~
{\buildrel e^2 \to 0 \over \sim}~
{1\over \#Z(G)} \int_{\CM} e^{\omega+ \epsilon \Theta}
 + \CO(e^{-c/e^2} )
}
$e^2=2 \pi^2 \epsilon$, $\alpha_2=(\rho,\rho)$, and
$c$ is a constant.
$\Theta\in H^4(\CM;\IQ)$ is - roughly- the characteristic
class obtained from
$c_2(\CQ)$ where  $\CQ\to \ST\times \CM^{\rm irr}$,
is the universal flat $G$-bundle.
\foot{Precise definitions are in \AtBoym.}
$\Theta$ is best thought of in terms of the
$\CG(P)$-equivariant cohomology of $\{A\in \CA(P): F(A)=0\}$.
In the Cartan model it is represented by
${1\over 8 \pi^2} \Tr \phi^2$.
The ``physical argument'' for \smllcpl\ proceeds by writing the
path integral as:
\eqn\physpf{
\eqalign{
Z(e^2, p,G) &\qquad\qquad\qquad\qquad\qquad\qquad\cr
= {1\over \vol \CG} &
\int d \phi d A d \psi exp\Biggl\{
\biggl[{i \over 4 \pi^2}
\int_\Sigma  \Tr( \phi F- \half \psi\wedge \psi)\biggr]
 +\biggl[\epsilon \int_\Sigma  \mu
{1\over 8 \pi^2} \Tr \phi^2 \biggr]\Biggr\} \cr}
}
where $\psi$ are the odd generators of the functions
on the superspace $\Pi T \CA$ and  $dA d \psi$
is the Berezin measure. This path integral
is the $t\to 0$ limit of  the partition function
of a cohomological field theory whose $Q$-exact
action is
$\Delta I = t Q \int \mu \Tr \psi^\alpha D_\alpha f $;
 $f=*F$,  $Q$ is the Cartan model
differential for $\CG$-equivariant cohomology
of $\CA$.  The partition function is $t$-independent
and localizes on the classical
solutions of Yang-Mills. A clever argument
 maps the theory at $t\to \infty$ to
``$D=2$ Donaldson theory''  and establishes the result.
{}From a mathematical perspective the first term in
the action of \physpf\ is the $\CG$-equivariant
extension of the moment map  on $\CA$,
the integral over $A,\psi$ defines an equivariant
differential form in $\O_\CG(\CA)$,
and the integral over $\phi$ defines equivariant
integration of such forms.
When the argument is applied to finite dimensional
integrals it leads to a rigorous result, namely the
nonabelian localization theorem for equivariant
integration of equivariant differential forms
\Witdgtr\jeffkir.

\newsec{ Large $N$ Limit: the Hilbert Space }

The large $N$ limit of \ymt\ amplitudes
is defined by taking  $N\to \infty$ asymptotics for
gauge group
$G=SU(N)$, holding $e^2a N\equiv \half \lambda$ fixed.
It is instructive to consider first the
Hilbert space of the theory. In the large
$N$ limit the statespace can be
described by the conformal field theory (CFT) of
free fermions
\refs{\MiPoetd,\Dosc,\Docft}.
Bosonization then provides the key
to a geometrical reformulation in
terms of coverings \GrTa.

Nonrelativistic free fermions on $S^1$
enter the theory since
class functions on $SU(N)$ can be mapped to
totally antisymmetric functions on the maximal torus.
The Slater determinants of $N$-body wavefunctions give
the numerators of the Weyl character formula. The
Fermi sea corresponds to the trivial representation
with one-body states $\psi(\theta) = e^{i n \theta}$
occupied for $\mid n \mid \leq \ha (N-1)$.
In the representation basis  the Hilbert space is:
$\CH_{SU(N)} = \oplus_{n\geq 0} \oplus_{Y\in \CY_n^{(N)}}
{\bf C}\cdot \mid R(Y)\rangle$;
$\CY_n$=the set of Young diagrams with $n$ boxes,
 $\CY_n^{(N)}$ is the subset of diagrams with $\leq N$ rows,
$R(Y)$ is the  $SU(N)$ representation  corresponding
to $Y\in \CY_n^{(N)}$.
The naive $N\to \infty$ limit of $\CH_{SU(N)}$ is
$\CH^+ = \oplus_{n\geq 0} \oplus_{Y\in \CY_n}
{\bf C}\cdot \mid Y \rangle$. The space $\CH^+$ is
related to the state space of a  c=1 CFT.
Excitations of energy
$\ll N$ around the
 Fermi level $n_F =  \ha (N-1)$ are described
using the zero-charge sector $\CH_{bc}^{Q=0}$ of a
``$\lambda=1/2$ $bc$ CFT''  \FMS, where $Q=\oint_{S^1} bc$.
The point of this reformulation is that one can
apply the well-known bosonization theorem
which relates the ``representation
basis'' to the  ``conjugacy class basis.''
Focusing on one Fermi level we
 define fermionic
oscillators $\{ b_n,c_m\}=\delta_{n+m,0}$,
a Heisenberg algebra
$[\alpha_n, \alpha_m] = n \delta_{n+m,0}$
related by $\alpha_n = \sum b_{n-m} c_m$,
and compare, at level $L_0=n$,  the fermionic basis:
$\mid Y(h_1, \dots h_s)\rangle =
c_{-h_1+1 - \ha}  \cdots c_{-h_s+s - \ha}
b_{-v_1 +1- \ha} \cdots b_{-v_s +s - \ha}\mid 0 \rangle
$ where
$Y(h_1, \dots h_s)\in \CY_n$ is a Young diagram with
row lengths $h_i$, with the bosonic basis
$ \mid \vec k \rangle
\equiv \prod_{j=1}^\infty (\alpha_{-j})^{k^j} \mid 0\rangle
$
where  $\vec k= (k_1, k_2, \dots) $
is a tuple of nonnegative integers, almost all $0$.
$\vec k$  specifies  a
partition of $n=\sum j k_j $ and  a conjugacy
class $C(\vec k) \subset S_n$.
The fermi/bose overlap is given by
the characters of the symmetric group representation $r(Y)$:
$\langle \vec k \mid Y\rangle = {1\over n!} \chi_{r(Y)} (C(\vec k))$.

When applying the above well-known technology to \ymt\ one
finds a crucial subtlety \GrTa:
$\CH^+$ is {\it not} the appropriate
limit for \ymt. At $N\le \infty$ there are
two Fermi levels $n_F=\pm \half (N-1)$; excitations
around these different levels are related to tensor
products of $N,\bar N$ representations, respectively.
In the large $N$ limit one must consider representations
occurring in the decomposition of
 tensor products $R\otimes \bar S$ where $R,S$ are
associated with Young diagrams with $n\ll N$ boxes,
and $\bar S$ is the conjugate representation. That is,
the correct limit   for \ymt\  is
 $\CH_{SU(N)} \to \CH^+\otimes \CH^- $.
The two $bc$ systems are naturally interpreted as
left- and right-moving sectors
of a  $c=1$  CFT.

Gross and Taylor provided an elegant
interpretation of the $N\to \infty$
\ymt\ Hilbert space
in terms of covering maps \GrTa.
The
one-body string Hilbert space is
identified with the group algebra
$\IC[\pi_1(S^1)]$. The state
$\mid \vec k \rangle\in \CH^+$ is identified with
a state in the Fock space of strings
defined by $k_j$
$j$-fold oriented coverings  $S^1\to S^1$.
The structure of the statespace
$ \CH^+\otimes \CH^- $ has a natural
geometrical interpretation in terms of
string states $\mid \vec k\rangle \otimes \mid \vec l\rangle $:
$\vec k, \vec l$
describe orientation preserving/reversing coverings.

\newsec{ $1/N$ Expansion of Amplitudes }

The $1/N$ expansion of \ymt\ has a very
interesting interpretation in terms of the
mathematics of covering spaces of
$\ST$. Heuristically the worldsheet
swept out by a $j$-fold cover $S^1\to S^1$
defines a $j$-fold cover of a cylinder by
a cylinder. Moreover, the Hamiltonian
$H=C_2$ is not diagonal in the string basis.
One finds
a cubic interaction term describing the
branched cover of a cylinder by a trinion
\MiPoetd\Dosc\Docft.

To state a more precise relation we
define the chiral partition function to be:
$Z^+(\lambda,p)\equiv \sum_{n\geq 0} \sum_{Y\in \CY_n }
(\dim R(Y))^{2-2p} e^{- \lambda C_2(R(Y))/(2N) }$.
$Z^+$ exists as an asymptotic expansion in $1/N$.
The $1/N$ expansion is related to
topological invariants of  Hurwitz spaces.
To define these let  $H(n,B,p,L)$
stand for the equivalence classes of
connected branched coverings of
$\Sigma_T$ of degree $n$,
branching number $B$ and  $L$ branch points.
If $\CC_L(\Sigma_T)\equiv \{(z_1,\ldots,z_L)\in
\Sigma_T^L\vert z_i \in \Sigma_T, z_i \ne z_j  \}/S_L$,
then
$H(n,B,p,L) \to \CC_L(\Sigma_T)$
is an unbranched cover
with discrete fiber above $S\in \CC_L$ given by
the equivalence classes of
homomorphisms $\pi_1(\ST - S, y_0)\to S_n$,
$y_0\notin S$ \Fulton.
Let $H(h,p)\equiv \amalg'_{n,B\geq 0} \amalg_{L=0}^B H( n,B,p,L)$
where the union on $n,B$ is taken consistent with
the Riemann-Hurwitz relation: $2h-2=n(2p-2)+B$.
We  define the orbifold Euler characters of
Hurwitz spaces by the formula
 $\chi_{\rm orb} \bigl\{ H( h,p)\bigr\}
\equiv \sum'_{n,B\geq 0} \sum_{L=0}^B
\chi (\CC_L(\Sigma_T) )\sum_{\pi_0(H( n,B,p,L))}
\vert Aut f\vert^{-1} $.

\noindent
{\bf Theorem 5.1} (\GrTa\ + \CMROLD). For $p>1$:
$$
Z^+(0,p)
{}~ {\buildrel N\to \infty\over \sim} ~
 \exp \bigl[  \sum_{h=0}^{\infty} \bigl({1\over N}\bigr)^{2h-2}
\chi_{\rm orb} \bigl\{ H( h,p)\bigr\} \bigr]
$$

\noindent
{\it Proof}:  In \GrTa\ Gross and Taylor used
 Schur-Weyl reciprocity to write $SU(N)$
representation-theoretic objects in terms of
 symmetric groups.
A key step was the introduction of an element of
the group algebra $\Omega_n = \sum_{v\in S_n}
\bigl({1\over N}\bigr)^{n-K_v} v\in \IC[S_n]$ where
$K_v$ is the number of cycles in $v$.
$\Omega_n$ is  invertible for
$N>n$ and satisfies: $(\dim R(Y))^m=
({N^n \dim r(Y) \over n!})^m
{\chi_{r(Y)}(\Omega_n^m)\over \dim r(Y)}$
for {\it all} integers $m$.  Gross and Taylor showed that:
\eqn\Zchir{
\eqalign{
Z^+(\lambda,p)
{}~& \sim ~
\sum_{n,i,t,h=0}^{\infty} e^{-n \lambda/2} (-1)^i {{ (
\lambda)^{i+t+h}} \over{i!t!h!}}
 \bigl({1\over N}\bigr)^{n(2p-2)+2h+ i+ 2t} {{n^h(n^2-n)^t }\over
{2^{t+h}
}}\cr
&
\sum_{p_1,\ldots,p_i \in T_{2,n}}
\sum_{s_1,t_1,\ldots,s_p,t_p\in S_n}
\biggl[ {1\over {n!}}\delta (p_1\cdots p_i \Omega_n^{2-2p}
\prod_{j=1}^p
s_jt_js_j^{-1}t_j^{-1}) \biggr]
.\cr}
 }
$T_{2,n}\subset S_n$ is the conjugacy class
of transpositions,
and $\delta$ acts on an element of the
group algebra by evaluation at $1$.
$\delta$ is nonvanishing when its argument
defines a
homomorphism $\psi: \pi_1(\ST-S,y_0) \to S_n$
for some subset $S\subset \ST$.
One now uses  Riemann's theorem
identifying equivalence classes of degree $n$
branched covers
branched at $S\in \CC_L(\Sigma_T)$ with equivalence
classes of homomorphisms
$\psi: \pi_1(\ST-S,y_0) \to S_n$ to interpret
\Zchir\ as a sum over branched covers. Expanding the
$\Omega^{-1}$ points to obtain the coefficients
of the $1/N$ expansion gives the
orbifold Euler characters of Hurwitz spaces
$\spadesuit$

The significance of this theorem is that it
relates \ymt\ to CohFT.
To see this note that branched
covers are related to holomorphic
maps.  Indeed, let
$\tilde \CM(\Sw, \ST)
 = \CC^\infty( \Sw, \ST )\times  {\rm Met} (\Sw)$;
$\Met(\Sw)$ is the space of
smooth Riemannian metrics on a 2-surface
$\Sw$ of genus $h$. The moduli space of holomorphic
maps is  $\Hol(\Sw,\ST) \equiv
\{(f,g)\in \tilde \CM(\Sw, \ST) : df \epsilon(g) = J df\}
/ \bigl( \Diff^+\sdtimes \Weyl(\Sw)\bigr)$;
$\epsilon(g)$ is the complex structure
on $\Sw$ inherited from $g$, $\Weyl(\Sw)$ is
the group of local conformal rescalings acting
on $\Met(\Sw)$.
The definition of orbifold
Euler character above is thus natural since the
action by $\Diff^+(\Sw)$ on $\tilde \CM$ has
fixed points at maps with automorphism:
$\chi_{\rm orb} (H(h,p))
= \chi_{\rm orb} (\Hol(\Sw,\ST))$.
As we explain in the next section,
CohFT partition functions are Euler
characters of vector bundles over moduli spaces.

Theorem 5.1 has been extended in many
directions to cover  other correlation functions
of \ymt\GrTa\CMROLD. The results are not yet
complete but are all in harmony with the identification
of \ymt\ as a CohFT.
 Wilson loop amplitudes are accounted for
by Hurwitz spaces for coverings
$\Sw \to \ST$ by manifolds with
boundary.
\foot{The $\SDiff(\ST)$ invariance of \ymt\
implies that Wilson loop averages define
infinitely many invariants of immersions
$S^1\to \ST$.}
A formula analogous to \Zchir\ for the full,  nonchiral
theory has been given in \GrTa. The proof is not
as rigorous as one might wish, but we do not
doubt the result. The analog of theorem 5.1
involves  ``coupled covers'' \CMROLD.
A  {\it coupled cover}
$f:\Sw\to \ST$ of
Riemann surfaces is a map such that on the normalization of
$N(\Sw)=N^+(\Sw)\amalg N^-(\Sw)$ along
the double points $\{Q_1,\dots Q_d\}$ of $\Sw$,
$N(f)= f^+\amalg f^-$ where
$f^+:  N^+(\Sw)\to \ST$ is holomorphic and
$f^-:  N^-(\Sw)\to \ST$ is antiholomorphic and
$\forall i $, ramification indices match:
 $Ram(f^+,Q_i^+)= Ram(f^-,Q_i^- )$.
One may define a  ``coupled Hurwitz space''
$\CC\CH(\Sw,\ST)$ along the lines of the purely
holomorphic theory. The $1/N$ expansion of the
partition function again generates
the Euler characters of $\CC\CH(\Sw,\ST)$,
 {\it if} coupled
covers with ramified double points
receive a weighting factor
$\prod_Q Ram(f^+,Q^+)$ in the calculation of
the Euler characteristic \CMROLD. Put differently, the
proper definition of ``coupled Hurwitz space''
involves a covering of the naive moduli space of
coupled covers. This point has {\it not} been
properly understood from the string
viewpoint.

Finally, the results need to be extended to
the case of nonzero area.
When $\lambda\not=0$ the expansion \Zchir\ and
its nonchiral analog have the form:
$
Z^+(\lambda, p) = \sum_{h\geq 0 } (1/N)^{ 2h-2} Z^+_{h,p} (\lambda)
 =\sum_{h\geq 0} (1/N)^{ 2h-2}\sum_{n} e^{-n\lambda/2}  Z^+_{n, h,p} (\lambda)
$.
For $p > 1$, $Z^+_{n,h,p} (\lambda) $ is polynomial in $\lambda$,
of degree at most $(2h-2)- n(2p-2)=B$. The string
interpretation described below shows that these polynomials
are related to intersection numbers in $H(h,p)$.
For $p=1$,  $Z^+_{h,1}(\lambda) $ are infinite
sums which can be calculated using
the relation to CFT described above \Docft. These
functions may be expressed in terms of Eisenstein
series and hence satisfy modular properties in
$\tau=i \lambda/(4\pi)$.
For example:  $Z_{1,1}^+
= e^{\lambda/48}\eta (i \lambda/(4\pi))$
($\eta$ is the Dedekind function)
\Grtalk.
The modularity in the coupling constant
might be an example of the phenomenon of
``S-duality'' which is currently under intensive
investigation in other theories.
For the case of a sphere: $ Z_{0,0} (\lambda)$  has finite radius of
convergence. At $\lambda= \pi^2$ there is a third
order phase transition (=discontinuity in the third
derivative of the free energy)  \DoKa. The existence
of such large $N$ phase transitions might present a serious
obstacle to a string formulation of higher-dimensional
Yang-Mills.

The $\lambda$-dependence of \Zchir\ has
been interpreted geometrically in
\GrTa. Contributions with $h,t>0$ are related to
degenerate $\Sw$. In the framework of
topological string theory the $h>0$ contributions are
probably related to the phenomenon of
bubbling \audin.

\newsec{ Cohomological Field Theory}

CohFT is the study of
intersection theory on moduli spaces using
quantum field theory. Reviews include
\refs{\cohoft,\bbrt,\Bl,\cmrrev}.  The
following discussion is a summary of the point
of view explained at length in \cmrrev.
In physics the
moduli spaces are presented as
$ \CM = \{ f \in \CC : D f =0 \} / \CG$
where $\CC$ is a  space of fields, $D$ is a differential
operator, and $\CG$ is a   group of local
transformations.  The action is an exact
form in a model for the $\CG$-equivariant cohomology
of a vector bundle over $\CC$.
The path integral localizes to the fixed points of the
differential $Q$ of equivariant cohomology.

More precisely, the following
construction of CohFT actions can be extracted
from the literature
\refs{\cohoft, \donaldson,\Witsm,\BaSiii,\BaSi,\AtJe,\stora,\Kal}.
We begin with the basic data:

1.)  $\CE\to \CC$, a  vector
bundle over field space which is
a sum of three factors:
$\CE=
\Pi\CE_{\rm loc} \oplus \CE_{\rm proj}
\oplus \Pi\CE_{\rm g.f. } $ (the $\Pi$ means
the fiber is considered odd).

2.) $\CG$-invariant metrics on $\CC$ and $\CE$.

3.) a  $\CG$-equivariant section $s: \CC\to \CE_{\rm loc}$,
a $\CG$-equivariant connection  $\nabla s= d s + \theta s \in
\O^1(\CC;\CE_{\rm loc}) $, and a
$\CG$-{\it nonequivariant}  section $\CF:\CC\to
\CE_{\rm g.f.}$ whose zeros determine
local cross-sections for $\CC\to \CC/\CG$.

The observables and action are best formulated
using the ``BRST model'' of $\CG$-equivariant
cohomology \refs{\stora,\Kal,\zuckerman}.
To any
Lie algebra $\lieg$ there is an associated
differential graded Lie algebra (DGLA)
$\lieg[\theta]\equiv \lieg \otimes \Lambda^*\theta$;
$\theta^2=0$, $\deg \theta=-1$, $\deg\lieg =0$,
$\p \theta=1$. Moreover, if $M$ is a superspace
with a $\lieg$-action then $\O^*(M)$ is a
differential graded $\lieg[\theta]$ module, with
$X\in \lieg \to \CL_X$, $X\otimes\theta\to \iota_X$.
In our case $\lieg\to \liebg$ and $M$ is the
total space of $\CE$.
The BRST complex is
$\hat\CE\equiv \Lambda^*\Sigma
(\liebg[\theta])^*\otimes \O^*(\CE)$
where $\Sigma$ is the suspension, increasing
grading by $1$. The differential on the
complex is
$Q=(d_\CE+ \p') + d_{C.E.} $ where
$\p'$ is dual to $\p$ and
$d_{C.E.}$ is the Chevalley-Eilenberg
differential for the DGLA $\liebg[\theta]$ acting on
$\O^*(\CE)$.   Physical
observables $\hat\CO_i$ are
 $Q$-cohomology classes of the ``basic'' ($\liebg$-relative)
 subcomplex and correspond to
basic forms $\CO_i\in \O^*(\CC)$ which descend
and restrict to cohomology
classes $\omega_i \in H^*(\CM)$.

The Lagrangian is $I = Q \Psi $, the
gauge fermion is a sum of three terms:
$\Psi  = \Psi_{\rm loc} + \Psi_{\rm proj}
+ \Psi_{\rm g.f.}$
for localization, projection, and gauge-fixing,
respectively. Denoting antighosts ($=$
generators of the functions on the fibers
of $\CE$) by $\rho+\theta \pi, \lambda+\theta \eta,
\bar c+\theta \bar \pi$, of degrees
$-1,-2,-1,$ respectively, and taking,
for definiteness, $\CE_{\rm g.f.}\mid_f=\CE_{\rm proj}\mid_f = \liebg$
we  have:
\eqn\ggfrms{\eqalign{
\Psi_{\rm loc}  &=
-i\langle \rho,s\rangle - (\rho,\theta \cdot \rho)_{\CE_{\rm loc}^*}
+{1\over 2}  (\rho,\pi)_{\CE_{\rm loc}^*}\cr
\Psi_{\rm proj}  & =i  (\lambda,C^\dagger)_{\liebg}\cr
\Psi_{\rm g.f.}  &=
\langle \bar c, \CF[A]\rangle -  (\bar c, \bar \pi)_{\liebg}\cr}
}
where $C^\dagger = (d R_f)^\dagger \in \O^1(\CC; \liebg)$,
is obtained, using the metrics, from the right $\CG$ action
through $f$,  $R_f:\CG\to \CC$.

The main result of the theory is a path-integral
representation for intersection numbers on $\CM$
as correlation functions in the cohomological
field theory:
\eqn\locfrm{
\int_{\hat\CE} \hat \mu e^{-I}
 \hat \CO_1  \cdots  \hat \CO_k
=
\int_{\CM=\CZ(s)/\CG} \chi\bigl[\cok(\bO)/\CG\bigr]
\wedge
\omega_1\wedge -\wedge \omega_k
}
where $\hat \mu$ is the Berezin measure
on $\hat\CE$
and
 $\bO = \nabla s \oplus C^\dagger
\in \O^1(\CC; \CV\oplus \liebg)$ is Fredholm
with $T\CM \cong \ker \bO/\CG$.
The argument for \locfrm\ may be sketched
as follows. The equations
$Df=0$ define the vanishing locus of a cross-section
$s(f)= Df \in \Gamma[\CE_{\rm loc} \to \CC]$.
Using the data of a metric and connection $\nabla$
on a vector bundle $E$, one constructs the
Mathai-Quillen representative $\Phi(E,\nabla)$ of
the Thom class of  $E$  \MaQu.  This construction
can be applied - formally -  in infinite dimensions
to write the Thom class for $\CE_{\rm loc}/\CG$.
When pulled back by a section $\bar s: \CC/\CG \to
\CE_{\rm loc}/\CG$, $\bar s^*(\Phi(\CE_{\rm loc}/\CG,\nabla))$
is Poincar\'e dual to the zero locus $\CZ(\bar s) = \CZ(s)/\CG$.
The natural connection on $\CE_{\rm loc}/\CG$ is
nonlocal in spacetime.
In order to find a useful field-theoretic representation
of the integral over $\CC/\CG$ one uses the
``projection gauge fermion''  $\Psi_{\rm proj}$ to
rewrite the expression as an integral over $\CC$.
Finally, one must divide by the volume of the
gauge group $\vol \CG$, necessitating the
introduction of $\Psi_{\rm g.f.}$.
The ``extra'' factor of $ \chi(\cok(\bO)/\CG)$
follows from a general topological
argument  \Winm\ or from a careful evaluation
of the measure near the $Q$-fixedpoints
$\spadesuit$

Two remarks are in order:
First, the factor $\chi(\cok(\bO)/\CG)$ is crucial in
studies of mirror symmetry \AsMo\wttnmirror\ and is
also crucial to the formulation of the \ymt\ string.
Second,
the formula \locfrm\  ignores
the (important) singularities in $\CM$.
We conclude with four examples:

{\bf 1}. ${\rm\underline{Donaldson\ Theory}}$
\refs{\donaldson,\BaSiii,\AtJe}:
$P\to M$ is a principal $G$-bundle over
a 4-fold $M$.
$\CC=\CA(P)$, $\CG=Aut(P)$,
$s(A)=F_+ \in \CE_{\rm loc}=\CA\times \O^{2,+}(M;\lieg)$.
$\cok\bO=\{ 0 \} $ (at irreducible connections).
Observables are $\int_\gamma \Phi^*(\xi)$;
$\gamma\in H_*(M),\xi\in H^*(BG)$,
$\Phi:(P\times\CA(P))/(G\times \CG(P))\to BG$ is
the classifying map of the $G$-bundle $\CQ\to \CA/\CG\times M$ of
Atiyah-Singer \atiyahsinger.
\locfrm\ becomes Witten's path integral representation
of the Donaldson polynomials.

{\bf 2}. $\underline{{\rm Topological}\
\sigma\ {\rm Model}}, T\sigma(X)$ \Witsm\BaSi:
$X=$ a compact, almost K\"ahler manifold
with almost complex structure $J$.
$\CC=\Map(\Sw,X)$.  $\Sw$ has complex
structure $\epsilon$ and
$s(f) = df + J df \epsilon\in
\CE_{\rm loc, f}= \Gamma(T^*\Sigma \otimes f^*TX)$.
Choosing a natural connection on
$\CE_{\rm loc}$ one finds
$\cok(\bO) \cong H^1(\Sigma, f^*(TX))$.
Observables are the Gromov-Witten classes:
$\int_\gamma \Phi^*(\xi)$;
$\gamma\in H_*(\Sw),\xi\in H^*(X)$,
$\Phi: \Sw\times \CC \to X$ is the universal map.

{\bf 3}. $ \underline{{\rm Topological\ String\ Theory}},  TS(X)$
\refs{\Witp,\VeVe,\DiVeVe}:
$X=$ compact, K\"ahler,
$\IF =(f,h)\in \CC=\tilde \CM(\Sw, X)
={\rm Map}(\Sw,X)\times \Met(\Sw) $,
$\CG=\Diff^+(\Sw)$,
$s(\IF) = (R(h)+1, df + J df \epsilon(h) ) $,
the first equation eliminating the Weyl
mode of the metric.
Observables are products of Gromov-Witten
classes and Mumford-Morita-Miller classes
on the moduli space of curves.

{\bf 4}. $ \underline{{\rm Euler}\  \sigma\ {\rm Model}}, \CE \sigma(X)$
\CMROLD\cmrrev: $X=$ compact, K\"ahler.
If, in $TS(X)$,  $\cok \bO= \{ 0\}$, $\CE \sigma(X)$
 computes the Euler character of
$\Hol(\Sw,X)$. The fieldspace
$\CC\to \tilde \CM(\Sw, X)$
 is a vector bundle with
fiber $\CE_{\rm loc}^*\oplus \liebg$. The section is:
$s(\IF, \hat \IF) = (s(\IF) , \bO^\dagger \hat \IF)$,
and, by construction, the partition function is:
$Z = \chi_{\rm orb} (\Hol(\Sw,X)) $, so $\CE \sigma(\ST)$
is the string
theory of (chiral) \ymt. The area dependence is obtained
by perturbing the action by
$\Delta I = \half \int f^* {\bf k}$ where ${\bf k}$ is
the K\"ahler class of $\ST$ (this is only
partially proven).
A similar construction reproduces
the nonchiral amplitudes but introduces an action
which is fourth-order in derivatives  and requires
 further investigation. An alternative
proposal for a string interpretation
of \ymt\ was made in \Hora. This approach
certainly  deserves further study.

\newsec{ Application and a Guess}

The original motivation for the program of
Gross was to find a
string interpretation of $YM_4$. Have we made
any progress towards this end? The answer is not
clear at present. We offer one suggestion here
in the form of a guess.
%
%\foot{existing
%evidence being too slim to dignify it with the
%name ``conjecture.''}
%

Combining \smllcpl\ with the $1/N$ asymptotics
of the \ymt\  partition function we expect
\foot{To make this statement rigorous one
must (a.) take care of the singularities in
$\CM$ and (b.) ensure that the corrections
$\sim \CO(e^{-2 N c/\lambda})$ from
\smllcpl\ are not overwhelmed by the
``entropy of unstable solutions''
\grssmatyt. The absence of phase
transitions as a function of $\lambda$ for
$G>1$
suggests that, for $G>1$, these terms
are indeed $\sim \CO(e^{- N c'})$ for
some constant $c'$. } an
intriguing relation between intersection theory
on $\CM(F=0,\ST)$ for $G=SU(N)$ and
the moduli spaces of holomorphic
maps $\Sw\to \ST$:
\eqn\combinging{
\eqalign{
\biggl \langle exp\biggl[ \omega +
{\lambda \over 4 \pi^2 N} \Theta
\biggr]\biggr \rangle
{}~
{\buildrel N\to \infty \over \sim}  &
\qquad\qquad\qquad\qquad\qquad \cr
C_N
\sum_{h\geq 0}  &\biggl({1\over N}\biggr)^{2h-2}
\sum_{d\geq 0}  e^{-\ha d  \lambda} P_d(\lambda)
\chi_{\rm orb}
(\CC\CH(\Sw,\ST,d))\cr}
}
$C_N=N e^{\alpha_1(2-2p)- \lambda \alpha_2/(2N)}$,
$\CC\CH(\Sw,\ST,d)$ is the coupled Hurwitz space
for maps of total degree $d$, and $P_d$ is a polynomial
with $P_d(0)=1$.

Now, the string theory of \ymt\
{\it does} have a natural extension to four-dimensional
target spaces:  $I= I(\CE \sigma(X))$ for
$X$ a compact  K\"ahler 4-fold.
Let
$e^\alpha$ be a basis of $H_2(X,\IZ)$ with Poincar\'e dual
basis ${\bf k}_\alpha$. The action may be
perturbed by
$ \Delta I = t^\alpha\int f^* {\bf k}_\alpha$.
Defining degrees $d_\alpha$ by:
$f(\Sigma) = \sum d_\alpha e^\alpha\in H_2(X,\IZ)$,
the partition function of the theory should have the
form
$Z(\CE \sigma(X))\sim \sum_{h\geq 0} \kappa^{2h-2}
\sum_{d_\alpha\geq 0} e^{-t^\alpha d_\alpha}
P_{d_\alpha}(t^\alpha) \chi(\CC\CH(\Sw, X; d_\alpha) )
$, more or less by construction,
where $\kappa$ is a string coupling constant and
$P_{d_\alpha}(t^\alpha)$ is a
polynomial whose value at zero is one.
Our guess is that a formula analogous to
\combinging\ holds in four dimensions, and that
 the asymptotic
expansion  of $Z(\CE \sigma(X))$
in $\kappa$ is closely related to the large $N$
asymptotics of intersection numbers of
the classes $\CO_2^{(2)}(e_\alpha)=\int_{e_\alpha} c_2(\CQ)$
on the moduli space of  antiselfdual instantons on $X$:
$\langle e^{r^\alpha \CO_2^{(2)}(e_\alpha)}
\rangle_{\CM_+(X;SU(N))}
$ where $\kappa\sim 1/N$ and $r^\alpha$
are analytic functions of
the  $t^\alpha$.

\bigskip
\centerline{\bf Acknowledgements}

The author is grateful to many colleagues for
essential discussions and collaboration on the
above issues; he thanks especially
S. Cordes, R. Dijkgraaf, M. Douglas,
E. Getzler, S. Ramgoolam, W. Taylor, and
G. Zuckerman. He thanks the theory group at
CERN for hospitality while the manuscript was
finished. He also thanks I. Frenkel for useful
comments on the manuscript.
This work is supported by DOE grants DE-AC02-76ER03075,
 DE-FG02-92ER25121,
and by a Presidential Young Investigator Award.

\listrefs

\bye